\documentstyle[aps,prb,twocolumn,eqsecnum,psfig]{revtex}
\input psfig

\def\etal{{\em et al.}}
\def\be{\begin{equation}}
\def\ee{\end{equation}}
\def\ber{\begin{eqnarray}}
\def\eer{\end{eqnarray}}
\def\bml{\begin{mathletters}}
\def\eml{\end{mathletters}}

\def\lav{\langle}
\def\rav{\rangle}

\def\tc{$T_c$} 
 
\def\pbtio{${\mbox{PbTiO}_3}\,$}
\def\pbzro{${\mbox{PbZrO}_3}\,$}
\def\batio{${\mbox{BaTiO}_3}\,$}

\def\khpo{${\mbox{KH}_2}{\mbox{PO}_4}\,$}
\def\nano{${\mbox{NaNO}_2}\,$}
\def\lb{\lambda_2}
\def\ld{\lambda_4}
\def\emax{$\epsilon_{\mbox{\footnotesize{max}}}$}
\def\dispvec{\xi}
\def\ddd{d}
\def\strain{\sigma}
\def\cbya{\frac{c}{a}}

\begin{document}
\preprint{TIFR/TH/1997-50}

\title{Theory of Size-Driven Transitions in 
Displacive and Order-Disorder Ferroelectrics}
\author{K. Sheshadri\cite{addshesh}, Rangan Lahiri\cite{addran} 
Pushan Ayyub\cite{addpushan} 
and Shobo Bhattacharya\cite{addshobo}}
\address{
Tata Institute of Fundamental Research, 
Homi Bhabha Road, Mumbai 400005, India}

\date{October 18, 1997}
\maketitle 

\begin{abstract}
We present a simple theory for structural 
transitions in displacive ferroelectrics of the perovskite type.
In our theory, the competition between the elastic energy cost for
the displacement of the homopolar ion from the centrosymmetric position,
and the energy gain due to a ferroelectric ordering of the dipoles formed
by the ionic displacements, leads naturally to a {\em first-order} transition
from a paraelectric to a ferroelectric phase. 
This transition takes place 
at a certain temperature $T_c(L)\/$ as the temperature is decreased and, at a
certain size $L_c (T)\/$ as the size of the system is increased. 
The transition temperature $T_c (L)\/$ is suppressed as 
the sample size is reduced, and vanishes for samples below a certain size.
For order-disorder ferroelectrics, our theory shows that 
the suppression of $T_c\/$ by a reduction in system size is not 
appreciable, a result that is borne out by experiments.

\end{abstract}

\section{introduction}
\label{sec:introduction}
Though the experimental study of finite size effects in 
ferroelectric materials has a long history \cite{anliker,kanzig}, 
the rapid development of advanced synthetic techniques has now 
made it possible to study different compounds in the form of phase-pure, 
ultrafine particles with a narrow size distribution. 
There is also a strong 
motivation for studying size-limited ferroelectric
systems in view of their current and potential applications as sensors, 
memory elements, nano-robotic and micro-electromechanical devices 
\cite{francombe}.

	A ferroelectric is termed `displacive' when the 
elementary dipoles strictly vanish in the paraelectric phase,
and `order-disorder'
when they are non-vanishing but thermally average 
out to zero in the paraelectric phase.
In recent years, the displacive ferroelectric transitions in \pbtio and \batio 
nanoparticles have been studied in detail.
The nature of the size effects observed in both 
systems is essentially similar. 
A decrease in the particle size 
causes monotonic reductions in (a) the transition temperature 
$T_c$, and (b) the tetragonal distortion 
of the unit cell which characterizes the ferroelectric phase. 
So, at a low 
enough particle size, the lattice tends to assume the high temperature, high 
symmetry, cubic paraelectric structure.

Uchino \etal \cite{uchino} have found that 
$T_c$ (in degrees Centigrade) falls with particle size $L$ (nm)  
following the relation
\be
T_c(L) = T_c(\infty ) - \frac{B}{(L-L_c)},
\label{eq:tcd}
\ee
where $T_c (\infty) = 128 ~(500), ~~ B=700 ~(588.5), \/$ 
and $ L_c = 110 ~(12.6)\/$ for \batio (\pbtio).

In a recent 
study of nanocrystalline \pbtio using dielectric, thermal and structural 
measurements \cite{soma95}, 
it was established that with decreasing particle size: 
(1) there is a monotonic decrease in the $T_c$, 
(2) the value of the peak dielectric constant (\emax) decreases,
(3) the ferroelectric transition becomes increasingly diffuse, and
(4) the crystallographic unit cell 
tends towards higher symmetry ($c/a \rightarrow 1$).
Qualitatively similar results have been obtained for \pbzro, a displacive 
{\em antiferroelectric} \cite{soma97}.

It is also instructive to compare finite size effects in displacive and 
order-disorder ferroelectrics. 
In electrically insulated samples of sub-micron \khpo,
a typical order-disorder system, the depolarization field appeared 
to prevent the stabilization of ferroelectric ordering below a critical size 
\cite{jacard}. However, a later study of \nano showed clearly that there was no 
suppression of $T_c$ down to 5 nm in samples suspended in either electrically 
insulated or conducting media \cite{marquardt}.
It is clear that the size driven effect is much weaker for 
order-disorder ferroelectrics.
This is a feature that our theory also captures (Section \ref{sec:od}).

	In recent years, there have been a few attempts to 
understand theoretically the nature of size effects in ferroelectrics. 
Using the 
phenomenological Landau-Devonshire theory, Zhong \etal \cite{zhongprb} have 
shown that the ferroelectric $T_c$ should decrease with decreasing size, 
ultimately leading to a size-driven phase transition from the ferroelectric to 
the paraelectric phase. Shih \etal \cite{shih} have considered the effect of 
incorporating the depolarization energy in the Landau free energy density.

In this paper, we restrict ourselves mainly to the description 
of quasi-free ferroelectric nanoparticles. 
The system is assumed to consist 
of loosely aggregated, un-clamped particles which are not electrically 
isolated. 
Under such circumstances, we can neglect the effects of external 
strain and depolarization. 
This approximates the experimental situation considered by 
Chattopadhyay \etal\/ \cite{soma95}.
Specifically, we have selected \pbtio as the model 
system, but the results should apply to other displacive-type systems as well. 

\pbtio is a classical displacive ferroelectric with a tetragonal 
perovskite structure ($a = 0.3899 {\mbox{nm}},~~ c = 0.4153 {\mbox{nm)}}$.
At room temperature, the Ti and O ions are 
displaced with respect to the Pb ions, parallel to the polar axis 
with displacements $\ddd_{\mbox{\footnotesize{Ti}}} = .018 {\mbox{nm}}, 
\ddd_{\mbox{O}} = .047{\mbox{nm}}$ 
\cite{shiranepepinsky}.
On increasing the temperature, the tetragonal ferroelectric undergoes a 
first order transition to a cubic paraelectric phase
($a = c = 0.396 {\mbox{nm}}$) at $T_c = 768K$.

The paper is organised as follows.
In section \ref{sec:model} we present our model and discuss the qualitative
physics of the phase transition.
Section \ref{sec:calculations} gives the details of the calculation,
and our results are presented in section \ref{sec:results}. 

\begin{figure}
\begin{center}
\leavevmode
\psfig{figure=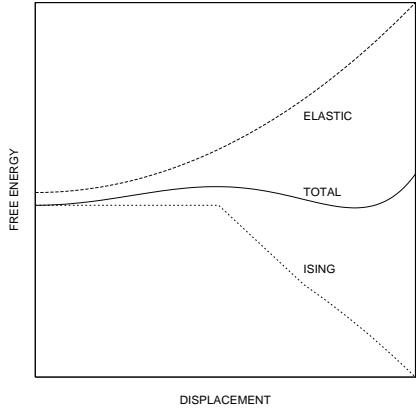,width=8cm}
\end{center}
\caption{
Schematic plots of the different contributions to the free energy
({\protect{\ref{eq:freeafterstrain}}}). The elastic part (dashed line) 
has a minimum at the origin. The second minimum is produced by the Ising 
part (dotted line) which
is constant up to a certain displacement and begins to decrease after 
that. This results in the full free energy with two minima (solid line).
}
\label{fig:freeenergy}
\end{figure}

In the picture adopted in the present paper, the transition takes place
as a result of competition between an ordering tendency of dipoles in
adjacent cells and the elastic energy cost associated with the displacement
of the atoms. Our main results are the following. We offer a simple
physical explanation (see Fig.\ref{fig:freeenergy})
for the first-order transition in displacive systems
driven by temperature and system size. We calculate the strain of the
unit cell and make comparison with experiments. While the temperature
dependence of the strain (Fig.\ref{fig:xvsT}) is not in very good agreement
with experiments,
the size dependence we obtain is in excellent agreement (Fig.\ref{fig:xvsL}). 
We show that for order-disorder ferroelectrics,
size effects are highly suppressed (section.\ref{sec:od}). We present 
the phase diagram for both kinds of ferroelectrics in the temperature - size 
plane in Fig.\ref{fig:TLpdiag}.
The paper ends with a brief discussion of certain features not captured
quantitatively in our theory, and prospects for future work 
(section \ref{sec:discussions}).

\section{The Model and the Physical Picture}
\label{sec:model}

A dipole moment is created in each unit cell
by the motion of the ionized atom at the centre of the
unit cell to an off-centre position.
This atom is known as the homopolar atom. In \pbtio and \batio for
instance, the Ti atom plays this role.
The displacement is accompanied by a distortion of the cubic unit cell
to a tetragonal one with sides $a\times a\times c$.
The order parameter  measured in the experiments \cite{soma95}, 
$\strain = c/a - 1$, obtained from powder x-ray scattering,
is a measure of the strain or distortion of the unit cell.

Clearly there is an energy cost at each site associated with the displacement 
\cite{fn:distortioninonedirection} $\dispvec_i$ of the homopolar atom.
In spite of this cost,
the system may find it profitable to undergo a
distortion accompanied by an off-center motion of the homopolar ion
if there is a negative energy contribution from the interaction energy
of dipoles when they are aligned.
We describe this energy phenomenologically by an interaction
$- J \dispvec_i \dispvec_j $ between nearest neighbours.
We refer to this term as the `Ising' term (see below).
For ferroelectrics, $J$ is positive, whereas a negative $J$ will
describe antiferroelectrics.
In the present work we focus mainly on ferroelectrics
(see however section \ref{sec:discussions}).

The effective Hamiltonian for the problem 
is thus the sum of an elastic part and an Ising part.
The form of the elastic part may be deduced from 
simple symmetry considerations.
The  cost of a displacement $\dispvec_i$ of the homopolar atom 
at the $i^{th}$ unit cell and 
an associated strain $\strain_i (\equiv c/a - 1)$ of the unit cell 
can be written as a
power-series expansion in $\dispvec_i$ and $\strain_i$, at each site.
Only even powers of $\dispvec_i$ are allowed in the expansion 
since the cost cannot depend on which way the atom moves.
There is no such restriction for the strain $\strain$, 
which is a scalar. 
We thus arrive at the Hamiltonian
\ber
H (\dispvec, \strain ) &=& \sum_{i} (\frac{1}{2} k_2 \dispvec_{i}^{2} + 
\frac{1}{4} k_4 \dispvec_{i}^{4} + a \strain_{i}^{2} 
-
b \strain_{i} \dispvec_{i}^{2})
\nonumber\\
& -& 
J\sum_{i,j} \dispvec_i \dispvec_j.
\label{eq:originalH}
\eer
There is no term proportional to $\strain_i$ alone, since when 
$\dispvec_i = 0$ for all $i$, 
the system should be in equilibrium with $\strain_i = 0$.

\begin{figure}
\begin{center}
\leavevmode
\psfig{figure=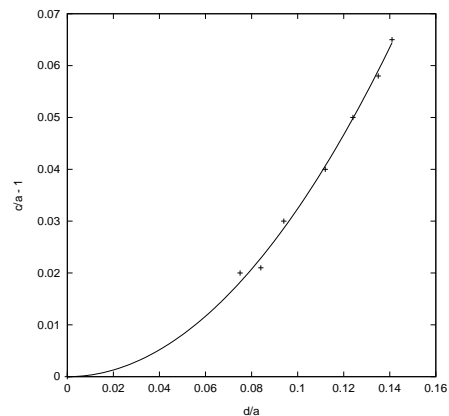,width=8cm}
\end{center}
\caption{
Experimental values of the strain ($\strain = \cbya -1$) 
plotted as a function of $\ddd/a$, the displacement $|\dispvec|$ of the
homopolar atom in units of lattice spacing {\protect{\cite{fn:dispfromP}}}.
The points correspond to measurements at different temperatures
below $T_c$, and the line is the best fit 
(Eq. {\protect{\ref{eq:strainvdispempirical}}}).
}
\label{fig:strainvdisp}
\end{figure}

An additional consideration has gone into the truncation of the power series.
It is expected on symmetry grounds that the thermally averaged
value $\lav \strain \rav$, of the strain will be an even function of the
thermally averaged displacement $\lav \dispvec \rav$, and for small distortions,
$\lav \strain \rav \sim {\lav \dispvec^2 \rav}$.
This is borne out extremely well from experimental data on \pbtio 
(Fig. \ref{fig:strainvdisp}).
From the data we find the empirical relation
\be
\lav \strain \rav \simeq A \frac{\lav | \dispvec|^2 \rav}{a^2}.
\label{eq:strainvdispempirical}
\ee
where $A = 3.24 $ for \pbtio.
The quadratic relation holds good for other oxides as well \cite{fn:dispfromP},
with different values of $A$.

We may conclude from the above discussion that ${\dispvec}^2$ and $\strain$
are of the same order of smallness.
The power series expansion in (\ref{eq:originalH}) thus retains all
terms upto ${\cal{O}}({\strain}^2)$ or ${\cal{O}}({\dispvec}^4)$.

Starting with the Hamiltonian (\ref{eq:originalH}), we can easily integrate
out the strain field $\strain$, to arrive at an effective free energy
in terms of the displacements,
\ber
F({\dispvec_i}) &=& \frac{1}{2} \lambda_2 \sum_{i}\dispvec_{i}^{2} 
+
\frac{1}{4} \lambda_4\sum_{i} \dispvec_{i}^{4}
\nonumber\\
&-&  J\sum_{<ij>} \dispvec_i \dispvec_j
\label{eq:freeafterstrain}
\eer
where 
$\lambda_2 = k_2 \/$ and $\lambda_4 = k_4 - \frac{b^2}{4a}\/$.

To get a physical picture of the temperature driven 
transition in bulk samples,
we introduce variables $\alpha_{i}$ and $\ddd_{i}$ corresponding to the
direction and magnitude of the displacement,
\be
\dispvec_i = \ddd_i \alpha_i,
\ee
where $\alpha_{i}$ ($= \pm 1$) is an Ising variable.
With this change of variables, the ferroelectric part of the
Hamiltonian (See Eq.(\ref{eq:freedispalpha})) is like a compressible
Ising term\cite{compressible}, since the coupling $\sim J \ddd^2\/$
depends on the displacement. At a temperature $T\/$, the
Ising part cannot be ordered (and hence the system cannot be a 
ferroelectric) if the coupling is less than $T\/$. This means
that the system can be a ferroelectric {\em only if} $\ddd^2 \/$
is greater than $\sim T/J\/$. Whether it actually is a ferroelectric
or a paraelectric depends on energetics. This explains
why we expect a first-order transition driven by temperature.

To capture this quantitatively, we may perform a partial trace over the 
$\alpha$ degree of freedom and evaluate a free energy functional of the
distortions $\ddd_i$ alone, given by
\be
e^{-F(\ddd)} =  Tr_{\alpha_i}  e^{- H(\alpha,\ddd)}.
\ee
The global minimum of $F({\ddd})$, determines the thermodynamic phase.
We shall see in the following section that 
this procedure automatically leads to a first order transition. 

The size induced transition is also easily understood within this picture.
For a system with $N = (L+1)^3\/$ sites, the elastic energy is $\sim N\/$ and
the Ising energy is $\sim Nz(L)\/$ where $z(L)\/$ is the average coordination
number of the cubic lattice of linear size $L\/$, given by

\ber
z(L) &=& \frac{1}{(L+1)^3} \left[ 6(L-1)^3 + 30(L-1)^2 \right.
\nonumber\\
&& + \left. 48(L-1) + 24 \right].
\label{eq:zofl}
\eer
For very large $L\/, z \rightarrow 6\/$ and the system, let us say, is 
ferroelectric provided $T < T_c\/$. If we reduce $L\/$, this reduces
$z(L)\/$, and the magnitude of the Ising part decreases as a result. 
This makes the ferroelectric phase unstable at small size and the 
system becomes a paraelectric.

In particular, at zero temperature, when there are no thermal fluctuations,
$\alpha_i$ is the same at all sites, and the free energy is simply
\ber
f(\ddd) = \frac{1}{2}(\lb - Jz(L))\ddd^2 + \frac{1}{4}\ld \ddd^4 \/, 
\eer
which describes a {\em second order} transition driven by size when
\ber
J z(L_c) = \lb 
\eer
at a certain critical size $L_c\/$.

\section{Calculations}
\label{sec:calculations}
In this section we present a mean-field theory for our model.
The effective Hamiltonian (\ref{eq:freeafterstrain})
expressed in terms of $\ddd_i$ and $\alpha_i$ is
\ber
F(\ddd,\alpha) &=& \frac{1}{2} \lambda_2 \sum_{i}\ddd_{i}^{2} 
+
\frac{1}{4} \lambda_4\sum_{i} \ddd_{i}^{4}
\nonumber\\
&-&  J\sum_{<ij>} \ddd_i \ddd_j \alpha_i \alpha_j.
\label{eq:freedispalpha}
\eer
Recall that $\ddd_i$ is the magnitude of the displacement of the
homopolar ion in the $i^{\mbox{th}}$ unit cell and $\alpha_i$ is its sign.
We now perform a trace over the $\alpha$ variables 
at the mean-field level, to get
\ber
F({\ddd_i}) &=& \frac{1}{2} \lambda_2 \sum_{i}\ddd_{i}^{2} 
+
\frac{1}{4} \lambda_4\sum_{i} \ddd_{i}^{4}
\nonumber\\
&+& \frac{1}{2} J\sum_{<ij>} m_i m_j \ddd_i \ddd_j
\nonumber\\
&-&
T\sum_{i}\ln 2\cosh (\frac{J \ddd_{i} \sum_{j}m_{j} \ddd_{j} }{T}).
\label{eq:free1}
\eer
The thermal average $m_{i}=\langle \alpha_{i} \rangle \/$ is determined by 
the self consistency equation
\begin{equation}
m_{i} = \tanh (\frac{J \ddd_{i} \sum_{j}m_{j} \ddd_{j} }{T}).
\label{eq:selfcon}
\end{equation}

In this paper, we resort to the approximation
$\ddd_{i}=\ddd\/$, i.e., we assume that
the homopolar atom displaces by the {\em same amount in all the unit cells}.
Therefore, the free energy per unit cell $f({\ddd_i})=F({\ddd_i})/N\/$ 
is given by (see Eq. (\ref{eq:free1}))

\begin{equation}
f(\ddd) = e(\ddd) + I(\ddd),
\label{eq:free2}
\end{equation}
where we have separated the elastic part 

\begin{equation}
e(\ddd) = \frac{1}{2} \lambda_2 \ddd^{2} + 
\frac{1}{4} \lambda_4 \ddd^{4},
\label{eq:elastic}
\end{equation}
and the Ising part
\begin{equation}
I(\ddd) = J \ddd^{2} \frac{1}{N} \sum_{<ij>} m_i m_j-
\frac{1}{N}\sum_{i} T\ln 2\cosh (\frac{J \ddd^{2}\sum_{j}m_{j} }{T}).
\label{eq:ising}
\end{equation}
For a bulk system ($L \rightarrow \infty$), the Eq.(\ref{eq:selfcon}) gives a
uniform solution $m_i = m\/$. We thus have

\begin{equation}
I(\ddd) = \frac{1}{2} J z m^2 \ddd^{2} -
T\ln 2\cosh (\frac{J z m \ddd^{2} }{T}).
\label{eq:isingbulk}
\end{equation}
where $z\/ (= 6)$ is the coordination number of the cubic lattice.

The schematic plots of $e(\ddd),I(\ddd)\/$ and $f(\ddd)\/$ are 
shown in Fig. \ref{fig:freeenergy}. 
It can be
seen that the elastic part is an increasing function of the displacement
with a minimum at $\ddd=0\/$,
whereas the Ising part is constant, equal to $-T \ln 2$, 
for $\ddd < \ddd_c = \sqrt {T/Jz}\/$ and 
decreases (i.e. increases in magnitude, being negative always)
for larger $\ddd\/$. 
Indeed, for $\ddd<\ddd_c, ~~ m=0\/$ 
(see Eq. (\ref{eq:selfcon})) and therefore $I(\ddd) = -T\ln 2\/$. 
As a result of this
behaviour of $I(\ddd)\/$, the free energy develops a second minimum 
at some $\ddd=\ddd_{0}\/$, which is found by solving the equation
$\partial f /\partial \ddd = 0\/$. 
For large $\ddd, ~~~I(\ddd)\/ \sim -\ddd^2\/$, so that 
the quartic term in $e(\ddd)\/$
dominates for large $\ddd\/$, resulting in a stable free energy function
as shown in Fig. \ref{fig:freeenergy}. It is this two-minima structure of 
the free energy
that gives rise to a first-order transition between the ferroelectric 
(the $\ddd=\ddd_{0}\/$ minimum) and paraelectric 
(the $\ddd=0\/$ minimum) phases.

Minimising the free-energy in Eq.(\ref{eq:free2}) with respect to 
$\ddd\/$, we get
\begin{equation}
\lambda_2 \ddd + \lambda_4 \ddd^3 - J z m^2 \ddd = 0,
\label{eq:condition}
\end{equation}
corresponding to the two minima, one at $\ddd=0\/$ and the other at

\begin{equation}
\ddd_0 = \sqrt\frac{Jzm^2 -\lambda_2 }{\lambda_4 }.
\label{eq:minima}
\end{equation}
At the first-order transition, the two minima coexist, i.e., $f(0)
=f(\ddd_0)\/$, or 
\begin{equation}
\frac{1}{2} \lambda_2 \ddd_{0}^{2} + 
\frac{1}{4} \lambda_4 \ddd_{0}^{4} + 
\frac{1}{2} J z m^2 \ddd_{0}^{2} -
T_c\ln \cosh (\frac{J zm^2 \ddd_{0}^{2} }{T_c}) = 0.
\label{eq:coexist}
\end{equation}

We need three conditions to fix 
the three parameters of our theory $\lb, \ld$, and $J\/$. 
The last two equations (\ref{eq:minima} and \ref{eq:coexist}) 
supply two of these conditions;
in these we use the experimental values of \tc and $\ddd_0$. 
The third condition is the equation $\lb = J z(L_c)\/$
derived at the end of the last section.

From experiments on \pbtio, we have $\ddd_0 = 0.299 \AA \/$, 
$T_c \simeq 768$K, and from the work of Zhong \etal \cite{zhongprb}, 
$L_c \simeq 10\/$.
Using these in the three equations above, we obtain 
$J=4.345909\times 10^4$K$\AA^{-2}$,
$\lb=2.3718\times 10^5$K$\AA^{-2}$, and 
$\ld=2.6529\times 10^5$K$\AA^{-4}$. 

Now it is not surprising
that since $Jz\ddd_0^2/T_c \simeq 30.27\/$, we get $m\simeq 1\/$
for all the sites (see Eq.(\ref{eq:selfcon})). This in fact turns out to be
the case also for a finite lattice, so that the equations (\ref{eq:minima})
and (\ref{eq:coexist}) can be used even in this case by replacing $z\/$ by
$z(L)\/$. 
The results from this completely agree with the results of
solving Eq.(\ref{eq:selfcon}) explicitly for a finite lattice with 
{\em open} boundary conditions, which we have performed numerically.

To obtain the phase diagram, we calculate $\ddd_0 (T) \/$ for each $L\/$
from equations (\ref{eq:minima}) and (\ref{eq:coexist}) (with $z(L)\/$
replacing $z\/$).

\section{results}
\label{sec:results}

\begin{figure}
\begin{center}
\leavevmode
\psfig{figure=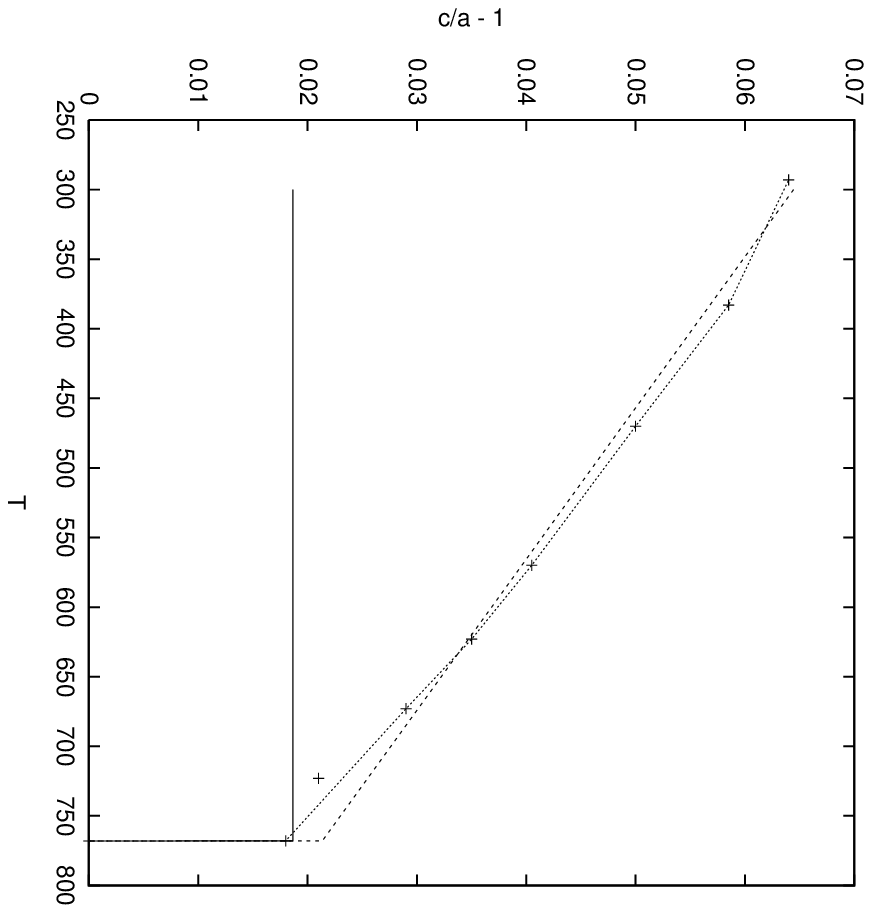,width=8cm,angle=-90}
\end{center}
\caption{
The strain plotted as a function of temperature ($T$) for 
bulk samples. The crosses are the experimental points and the solid line
is from the theory. The dotted line is obtained by assuming a weak
linear temperature dependence for $\lambda_2\/$, and the dashed line
is by assuming a temperature dependent $\lambda_4\/$ 
(Fig.{\protect{\ref{fig:lambda4}}}).
}
\label{fig:xvsT}
\end{figure}

Fig.\ref{fig:xvsT}
shows our result for the strain as a function of temperature for the
bulk case (solid line); the experimental data\cite{fn:dispfromP} 
are also shown (crosses).
The lack of quantitative agreement is ascribed to the fact that in our
mean-field theory, the ferroelectric is completely saturated, i.e., 
$m\simeq 1\/$ at all $T<T_c\/$. 

We can make up for this disagreement in two ways. One possibility is to assume
a weak temperature dependence for $\lambda_2\/$, of the form $A T + B\/$;
For $B=5.45J, ~~~ A=5.5 \times 10^{-4}J, ~~~ \lambda_4 = 
1.23 J \AA^{-2}\/$  and $J=1.65\times 10^{5}$K$\AA^{-2}$
($T\/$ measured in degrees Kelvin)
this leads to the dotted line shown in Fig.\ref{fig:xvsT}. We find that
while this makes the order parameter temperature dependent, it also makes
it $J$-independent and therefore size-independent, though 
the size-driven transition still remains. 

\begin{figure}
\begin{center}
\leavevmode
\psfig{figure=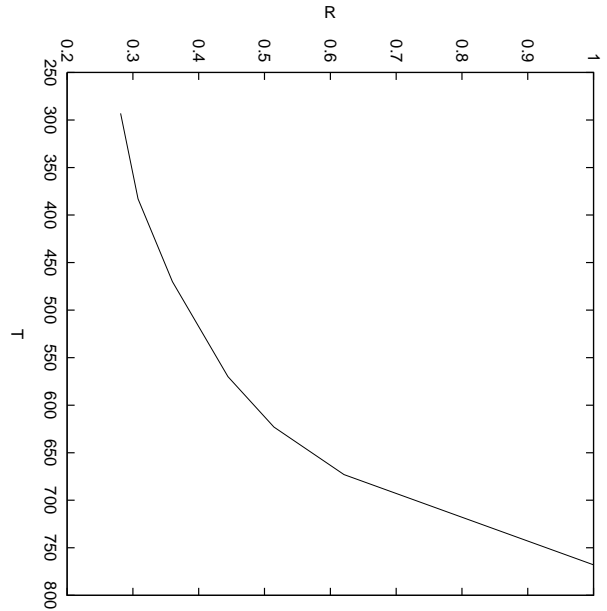,width=8cm,angle=-90}
\end{center}
\caption{
The ratio $R=\lambda_4 (T)/\lambda_4 (T_c)\/$ plotted as a function
of temperature. This dependence leads to the dashed line shown in
Fig.\protect{\ref{fig:xvsT}}.
}
\label{fig:lambda4}
\end{figure}

Alternatively, we can assume a temperature dependence
for $\lambda_4\/$ which can be calculated by requiring that
the order parameter agree with experiment. This turns out to be a rather
strong temperature dependence for $\lambda_4\/$ which we show in 
Fig.\ref{fig:lambda4}. 
Our subsequent results are obtained by assuming that the
parameters $\lambda_2\/$ and $\lambda_4\/$ are independent of temperature.

\begin{figure}
\begin{center}
\leavevmode
\psfig{figure=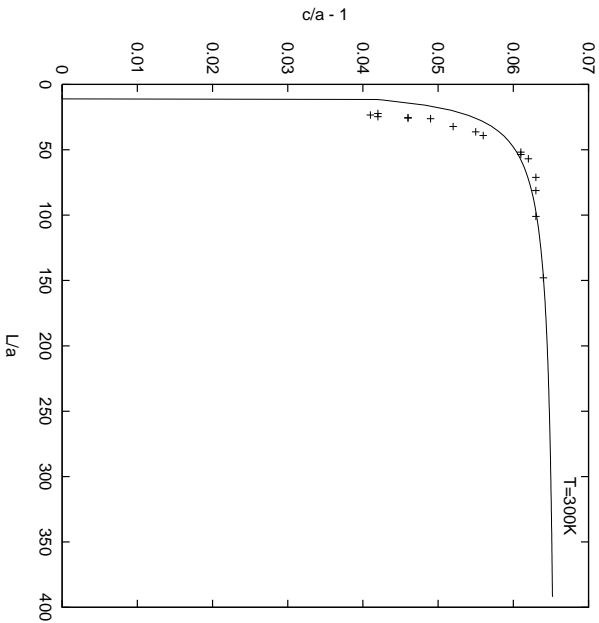,width=8cm,angle=-90}
\end{center}
\caption{
The strain in the tetragonal phase ($\cbya - 1$) as a function of reduced
system size at room temperature.  The solid line shows the strain 
calculated from our theory (see text). The crosses are experimental points 
{\protect{\cite{soma95}}}. 
}
\label{fig:xvsL}
\end{figure}

We have shown in Fig.\ref{fig:xvsL} a plot of strain calculated as a 
function of system size at a temperature of 300 K, along with the
corresponding experimental data\cite{soma95}. In this graph, we have 
normalized the calculated strain in a way so as to match
the experimental value for the bulk system at this temperature.
We obtain excellent agreement with experiment for the
size dependence of the strain.

\begin{figure}
\begin{center}
\leavevmode
\psfig{figure=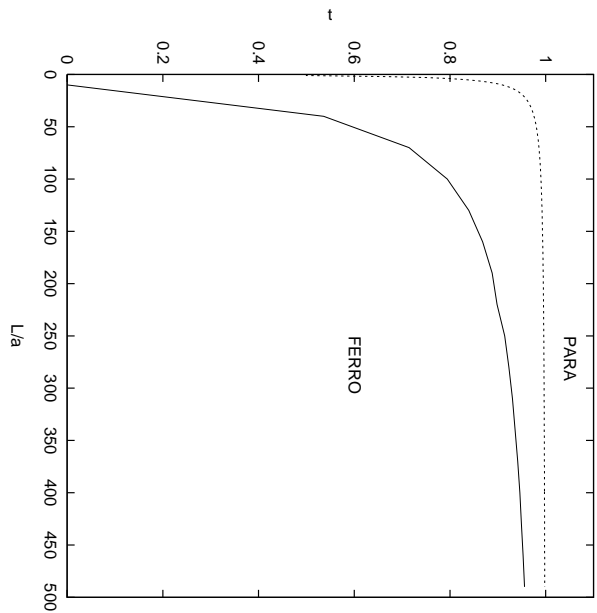,width=8cm,angle=-90}
\end{center}
\caption{
The phase diagram of ferroelectric oxides in the temperature-size 
plane obtained within our theory. We have plotted the reduced temperature
$t = T_c (L)/T_c (\infty) \/$ against the reduced particle size.
The solid line is for displacive
systems and is a first-order line separating the tetragonal ferroelectric
phase from the cubic paraelectric phase. It can be seen that there is
an appreciable suppression of $T_c\/$ as the system size is reduced, and 
$T_c\/$ is zero below a certain size. The dotted line is for 
order-disorder systems and is a line of second-order transitions separating
the low-temperature ferroelectric phase from the high temperature 
paraelectric phase. The transition temperature $T_c\/$ remains practically
equal to its bulk value down to very small system sizes, and the
size effects are highly suppressed in this case. In particular, the
system remains in the ordered phase below $T_c/2\/$ for all sizes.
}
\label{fig:TLpdiag}
\end{figure}

Fig.\ref{fig:TLpdiag} shows our 
phase diagram in the temperature - size plane, obtained by calculating
strain. The phase boundary for displacive systems 
(solid line) is a line of first-order transitions 
for all $T>0\/$, and the size-driven transition at $T=0\/$ is continuous,
as noted above. It can be seen from the phase diagram that our theory
predicts a suppression of $T_c\/$ as the system size is reduced, in
agreement with experiments on displacive systems and also other theoretical
studies on these systems.

\section{order-disorder ferroelectrics}
\label{sec:od}
For order-disorder ferroelectrics, the local potential experienced by the 
homopolar atom has a double well structure\cite{jona}, i.e., 
there is a local maximum at $\ddd =0\/$ and two minima at $\ddd =
\pm \ddd_0\/$. This corresponds to
$\lb < 0\/$ within our approach. 

In this case, it is easy to see that size effects are highly suppressed. 
Firstly,
we note that the Ising term can never lead to a minimum at $\ddd =0\/$
at any finite temperature. Therefore an increase in temperature can destroy
the ferroelectric order via a continuous transition. 
The distortion continues to have the value given in Eq.(\ref{eq:minima}),
but $m = \langle \alpha \rangle \/$ vanishes when $Jz(L)\ddd_{0}^{2}/T=1\/$.
The last condition can be met either by changing $T\/$ or by changing
system size. For the bulk system (z = 6), this is satisfied for $T=T_c\/$.
We can therefore write $z(L)/6 \simeq T/T_c\/$, where $L\/$ is the size at
which there is a size-driven transition at temperature $T\/$. 
Note that when expressed in terms of the reduced transition temperature
$t = T_c (L)/T_c (\infty)\/$ the above equation (which describes the
boundary between the ferroelectric and paraelectric phases) is independent
of our model parameters. If $T/T_c\/$
is about 0.9 (say), then we have to go to such small systems as $L\sim 10\/$
(see Eq.(\ref{eq:zofl})) to see size effects. 

Typically, the lattice
spacings in these systems are $\sim 4\AA\/$, so we have to get samples
of size $\sim 4\/$nm, to have observable 
size effects, even at so high a temperature as $0.9 T_c\/$. At lower 
temperatures the critical sizes are even smaller. In particular, for
$T<0.5T_c\/$, we do not expect any size-driven transition, since $z(L)\/$
cannot be less than 3. 
The experiment on NaNO$_{2}$\cite{marquardt} that we referred to in 
Sec.\ref{sec:introduction} reports that $T_c\/$ is not suppressed 
down to sizes of 5 nm.

We have shown the phase diagram for order-disorder ferroelectrics also
in Fig.\ref{fig:TLpdiag}, and the difference between the two kinds of
ferroelectrics is very instructive. We attribute the suppression of
size effects in the case of order-disorder ferroelectrics to the absence
of a structural transition accompanying the ferroelectric transition.
Indeed, at any temperature, one has to go to much smaller samples to
meet the requirement $Jz(L)\ddd_{0}^{2}/T=1\/$ (which is the transition
condition for order-disorder systems) than the co-existence
condition Eq.(\ref{eq:coexist}) (which is the transition condition for 
displacive systems).

\section{Discussions }
\label{sec:discussions}

Although our model gives a useful physical picture of both temperature
and size driven transitions, it is rather too simple to provide perfect
quantitative agreement.
We have made several simplifying assumptions, the most significant being
that the displacement has the same magnitude throughout the sample.
This is not necessarily true, especially for small systems where the
displacement at the surface could be different from that it is in the bulk.
This, and the subsequent neglect of thermal fluctuations, turn out
to be rather drastic approximations.
The strongly first-order character of the transition (i.e. the 
very weak, almost nonexistent temperature dependence of the
order parameter) is probably an artifact of our mean-field approximation. 

An interesting question, which cannot be answered
within the framework of our theory, is whether the transition remains
first-order beyond mean-field theory. Further, since our theory is an 
effective single-site theory, it cannot capture capture the phenomenon
of mode softening that is observed in these systems\cite{soft}.

We do not have the effect of depolarization field in our theory.
The depolarization field,
under certain circumstances, can lead to better quantitative agreement with 
experiment \cite{shih}.
However, in the experimental situation that we are describing \cite{soma95},
the effect of the depolarization field should be negligible.
While we have not made explicit the role of
long range interactions between dipoles, this will simply lead to
a reparametrisation of $J\/$ within the simple version of our theory
presented in this paper. 

The mean-field theory that we have attempted on the Ising part of the
Hamiltonian consists in reducing the lattice problem 
to an effective single-site problem.
This led to a very weakly temperature order parameter as we have seen. 
We have also performed calculations based on 
the Bethe-Peierls approximation\cite{huang}, which takes
into account short range correlations between dipoles. However, this did not
lead to any better temperature dependence for the order parameter
\cite{unpublished}.

There are certain qualitative merits of our theory that are worth emphasizing.  
For example, 
our theory can be easily extended to include antiferroelectric oxides,
simply by having a negative $J$.
The ordered phase will have a nonzero value for an antiferroelectric order 
parameter such as sublattice magnetization $\alpha_{\mbox{s}}$.
The molecular field at any site $i$, due to the nearest neighbours,
will be opposite in sign to the order parameter at that site,
so that the product of the molecular field and $J$ will have the
same sign as for the ferroelectric case.
The theory as worked out above in terms of displacements will go 
through without any further change, and the results will be identical
to that for ferroelectrics, with $J$ replaced by $|J|$.
Recent experimental studies of size-driven transitions in antiferroelectric
materials \cite{soma97} report results very similar to those
as ferroelectrics, confirming our expectation.

We can explain the form of Eq.(\ref{eq:tcd}) for the size-driven transitions
quite simply in our theory. For the size-driven transition, we get from
Eq.(\ref{eq:coexist}), upon
replacing $\lb, \ld,$ and $J\/$ by their values determined from bulk
data, an equation of the form $\ln \cosh (z(L)/T_c) = z(L)/T_c - A$, for some
constant $A\/$. It is clear that this equation will have a solution of
the form (\ref{eq:tcd}) for large $L\/$ since $z(L)\/$ is given by 
Eq.(\ref{eq:zofl}). The actual numbers appearing in Eq.(\ref{eq:tcd})
will of course depend on the system we address.

Our theory has three fitting parameters, whereas the phenomenological
Landau theories\cite{zhongprb,shih} have about twice as many.
This circumstance is clearly because our theory is based on a 
microscopic model which identifies the different contributions
to the system free energy as being due to the elastic and ferroelectric
parts. As a result, we are also able to describe the {\em structural}
transition in the displacive systems in addition to the ferroelectric
transition, to which it is related. The chief qualitative merit
of our theory is that the first-order transition appears in a very
natural way, with minimal assumptions about the form of the elastic
part of the Hamiltonian. With the lowest order nonlinear elastic term
(i.e. fourth order), we are able to describe the first-order
transition. This is an essential difference with earlier approaches
based on a Landau expansion of the free energy.

Finally, we have shown that our theory can describe size-driven transitions
in order-disorder systems as well as displacive ones. Our simple estimate
for the critical size at which size effects become important 
in these systems is in excellent agreement with data on 
NaNO$_2$\cite{marquardt}.
Our approach is superior to the Landau theory for size-driven
transitions \cite {zhongprb} 
since size effects in order-disorder systems can be addressed in our framework.

In conclusion, we have presented a simple unified picture for structural
transitions in displacive  and order-disorder ferroelectric oxides, 
which captures the physics of both temperature-induced transitions in 
the bulk and size induced transitions in nanoparticles.
The size effects are understood in a simple way as being the result
of smaller coordination number near the surface of the particle.
Our mean-field approximation gives qualitatively good results, except
for the temperature dependence of the order parameter.
Future work will center around incorporating the effect of thermal
fluctuations and getting closer agreement with experiments.

\acknowledgments

We would like to thank Mustansir Barma 
whose suggestions and critical remarks have gone a long way in helping
us understand the intricacies of the problem.
We thank Deepak Dhar for being the perfect bouncing board.
We have also benefited from discussions with 
Soma Chattopadhyay, Chandan Dasgupta, Rahul Pandit, T.V. Ramakrishnan,
Mohit Randeria, Madan Rao and Nandini Trivedi.
We thank Mustansir Barma, Deepak Dhar, and Arun Paramekanti
for their comments on the manuscript.  SB acknowledges hospitality of TIFR.


\begin{references}
\bibitem{fn:distortioninonedirection}
We have assumed above that the the the distortion of the unit cell
takes place along the same axis all through the sample, i.e., 
all the unit cells have their c axis along, say, the z direction.
This is reasonable, since there would be enormous elastic costs involved
in having adjacent unit cells distended along different directions.
Accordingly, the displacement can only be along the $z$ axis, but
it can be either up or down.
\end{references}
\end{document}